\documentclass[12pt]{article}
\def\nabstar#1{\nabla\kern-0.5pt\smash{\raise 4.5pt\hbox{$\ast$}}
               \kern-4.5pt_{#1}}

\def\drvstar#1{\partial\kern-0.5pt\smash{\raise 4.5pt\hbox{$\ast$}}
               \kern-5.0pt_{#1}}

\def\newline{\relax\ifhmode\null\hfil\break\else\nonhmodeerr@\newline\fi}
\def\frac#1#2{{#1\over#2}}
\def\text#1{{\hbox{\rm #1}}}

\newcommand{\beq}{\begin{equation}}
\newcommand{\eeq}{\end{equation}}
\newcommand{\bea}{\begin{eqnarray}}
\newcommand{\eea}{\end{eqnarray}}

\def\BE{\begin{equation}}
\def\EE{\end{equation}}
\def\BA{\begin{eqnarray}}
\def\EA{\end{eqnarray}}
\def\BAN{\begin{eqnarray*}}
\def\EAN{\end{eqnarray*}}

\def\nn{\nonumber\\}

\def\gm5{\gamma_5}

\input epsf.sty
\newdimen\psfigsize
\def\psfigure#1 #2 #3 #4 #5{
    \begin{figure}[tbh]
      \begin{center}
      \vbox{
        \null\vskip-0.2in\hskip#2
        \epsfxsize=#1
        \epsfbox{#4}
        \vskip -0.3in
        \caption {#5 \label{#3}}
        \vskip 0.0 true in plus 0.3 true in
      }
      \end{center}
   \end{figure}
}
\begin{document}
\thispagestyle{empty}
\begin{flushright}
NTUTH-02-505G \\
UW-PT-02-22 \\
September 2002
\end{flushright}
\bigskip\bigskip\bigskip
\vskip 2.5truecm
\begin{center}
{\LARGE {Optimal lattice domain-wall fermions}}
\end{center}
\vskip 1.0truecm
\centerline{Ting-Wai Chiu}
\vskip5mm
\centerline{Department of Physics, University of Washington}
\centerline{Seattle, WA 98195-1560, USA}
\smallskip
\centerline{and}
\smallskip
\centerline{Department of Physics, National Taiwan University}
\centerline{Taipei, Taiwan 106, Taiwan.}
\centerline{\it E-mail : twchiu@phys.ntu.edu.tw}
\vskip 1cm
\bigskip \nopagebreak \begin{abstract}
\noindent

I show that the conventional formulations of lattice domain-wall fermion
with any finite $ N_s $ (in the fifth dimension) do not preserve the
chiral symmetry optimally and propose a new action which preserves
the chiral symmetry optimally for any finite $ N_s $.



\end{abstract}
\vskip 1.5cm

\newpage\setcounter{page}1

Currently, there are many interesting physics issues
in supersymmetry, supergravity and superstring theory
that require nonperturbative (numerical) studies, 
in addition to those longstanding ones in QCD.    
A viable approach is to formulate these theories on a
spacetime lattice with domain-wall fermions (DWF).
The basic idea of DWF \cite{Rubakov:bb,Callan:sa}
is to use an infinite set of coupled Dirac fermion fields
[ $ \psi_s (x), s \in (-\infty, \infty) $ ] with masses
behaving like a step function $ m(s) = m \theta(s) $ such that Weyl 
fermion states can arise as zeromodes bound to the mass defect 
at $ s = 0 $. However, if one uses a compact set of masses, then
the boundary conditions of the mass (step) function must
lead to the occurrence of both left-handed and right-handed
chiral fermion fields, i.e., a vector-like theory.
For lattice QCD with DWF \cite{Kaplan:1992bt}, in practice,
one can only use a finite number ($ N_s $) of
lattice Dirac fermion fields to set up the domain wall, thus the
chiral symmetry of the light fermion field is broken.
Now the relevant question is how to construct
the couplings between these $ N_s $ lattice Dirac fermion fields
such that the exact chiral symmetry can be preserved optimally,
or in other words, the residual mass of the quark field
is the minimal.
Similarly, in numerical studies of $ {\cal N}=1 $ supersymmetric
$ SU(n) $ Yang-Mills theory with $ N_s $ domain-wall fermions,
it is vital to implement the chiral symmetry of the gaugino optimally
such that the supersymmetry can {\it most} easily emerge as
an ``accidental" continuum symmetry on the lattice.
In this paper, I discuss how to preserve the chiral symmetry of the light 
fermion fields optimally, for any finite $ N_s $ (in the fifth dimension).

First, we examine the domain-wall fermion action\footnote{In this paper,
we suppress the lattice spacings ($ a $ and $ a_5 $), as well as
the Dirac and color indices, which can be easily restored.
Also, it is understood that one can replace $ D_w $ with its improved
lattice Dirac operator, e.g., including the clover-like terms.}
with open boundary conditions \cite{Shamir:1993zy}, in the context
of lattice QCD,
\bea
\label{eq:shamir}
{\cal A}_{dwf} = \sum_{s,s'=1}^{N_s} \sum_{x,x'}
\bar\psi(x,s)
[ D_w(x,x') \delta_{s,s'} + \delta_{x,x'}  D_5(s,s') ] \psi(x',s')
\eea
where $ D_w $ is the 4D Wilson-Dirac operator with a negative
parameter $ -m_0 $
\BAN
D_w &=& \sum_{\mu=1}^4 \gamma_\mu t_\mu + W - m_0,
                       \hspace{6mm} m_0 \in (0,2) \\
t_\mu(x,x') &=& \frac{1}{2}[   U_\mu(x) \delta_{x',x+\mu}
                              - U_\mu^{\dagger}(x') \delta_{x',x-\mu} ] \\
W(x,x') &=& \sum_{\mu=1}^4 \frac{1}{2}[ 2 \delta_{x,x'}
                          - U_\mu(x) \delta_{x',x+\mu}
                          - U_\mu^{\dagger}(x') \delta_{x',x-\mu} ]
\EAN
and
\BAN
\label{eq:D5}
D_5 (s,s') 
= \delta_{s,s'} - P_{-} \delta_{s',s+1} - P_{+} \delta_{s',s-1},
\hspace{6mm}  P_{\pm} = \frac{1}{2} ( 1 \pm \gamma_5 ) .
\EAN
The boundary conditions are fixed by 
$ P_{+} \psi(x,0) = P_{-} \psi(x, N_s + 1) = 0 $. 
The quark fields coupling to physical hadrons can be constructed
from the left and right boundary modes
\BAN
q(x) &=& P_{-} \psi(x,1) + P_{+} \psi(x, N_s ) \\
\bar{q}(x) &=& \bar\psi(x,1) P_{+} + \bar\psi(x, N_s) P_{-}
\EAN
Then the quark propagator in a background gauge field can be evaluated
\cite{Neuberger:1997bg,Kikukawa:1999sy} as
\bea
\label{eq:propagator}
\left< q(x) \bar{q}(y) \right>
= \frac{1 - \gamma_5 S}{1 + \gamma_5 S}
\eea
where
\bea
\label{eq:S}
S &=& \frac{1-T}{1+T} \\
\label{eq:T}
T &=& \frac{(1-H)^{N_s}}{(1+H)^{N_s}} \\
\label{eq:h}
H &=& \gamma_5 \frac{D_w}{2 + D_w}
\eea

In the limit $ N_s \to \infty $, $ S \to H/\sqrt{H^2} $ 
(the sign function of $ H $), 
then the quark propagator (\ref{eq:propagator}) is chirally symmetric.

However, for any finite $ N_s $, (\ref{eq:propagator}) does not break
the chiral symmetry in the minimal way. In other words,
$ S $ (\ref{eq:S}) is {\it not} the optimal approximation
for the sign function of $ H $. This can be shown as follows.

First, we rewrite (\ref{eq:S}) as the partial fraction
\bea
\label{eq:S_polar}
S =  \left\{ \begin{array}{ll}
             H \left( \frac{1}{N_s} +
             \frac{2}{N_s} \sum_{l=1}^{n} \frac{b_l}{H^2 + d_l} \right)
             \equiv H R^{(n,n)}(H^2),
             &  \ N_s = 2n+1 \mbox{ (odd) } \nn
             H \
             \frac{2}{N_s} \sum_{l=1}^{n} \frac{b_l}{H^2 + d_l}
             \equiv H R^{(n-1,n)}(H^2),
             &  \ N_s = 2n  \mbox{ (even) }  \nn
             \end{array} \right. \\
\eea
where
\BAN
\label{eq:polar_bd}
b_l = \sec^2 \left[ \frac{\pi}{N_s} \left( l-\frac{1}{2} \right) \right],
\hspace{4mm}
d_l = \tan^2 \left[ \frac{\pi}{N_s} \left( l-\frac{1}{2} \right) \right] \ .
\EAN
Here the symbol $ r^{(n,m)}(x) $ denotes an irreducible rational
polynomial of the form
\BAN
\label{eq:rnm}
r^{(n,m)}(x)=
\frac{ p_{n} x^{n} + p_{n-1} x^{n-1} + \cdots + p_0 }
     { q_{m} x^{m} + q_{m-1} x^{m-1} + \cdots + q_0 } \ ,
     \ ( m \ge n, \ p_i, q_i > 0 )
\EAN

Note that the coefficients $ b_l $ and $ d_l $ in (\ref{eq:S_polar}) are
independent of (the ratio of the maximum and the minimum of) the
eigenvalues of $ H^2 $. As it will become clear later, this feature
already rules out the possibility that $ R^{(n-1,n)}(H^2) $ or
$ R^{(n,n)}(H^2) $
can be the optimal rational approximation of $ (H^2)^{-1/2} $.

According to de la Vall\'{e}e-Poussin's theorem
and Chebycheff's theorem \cite{Akhiezer:1992},
the necessary and sufficient condition for $ r^{(n,m)}(x) $ to be
the optimal rational polynomial of the inverse square root function
$ x^{-1/2} $, $ 0 < x_{min} \le x \le  x_{max} $ is that
$ \delta(x) \equiv 1 - \sqrt{x} \ r^{(n,m)}(x) $ has $ n + m + 2 $
alternate change of sign in the interval $ [ x_{min}, x_{max} ] $,
and attains its maxima and minima (all with equal magnitude), say,
\BAN
\delta(x) =  -\Delta, +\Delta, \cdots, (-1)^{n+m+2} \Delta
\EAN
at consecutive points ($ x_i, i=1,\cdots, n+m+2 $)
\BAN
x_{min} = x_1 < x_2 < \cdots < x_{n+m+2} = x_{max}\ .
\EAN

Now, for $ N_s = 2n $, $ \delta(x)=1- x R^{(n-1,n)}(x^2) $  is
non-negative for $ x > 0 $.
Thus $ \delta(x) $ does not have
any alternate change of sign for any intervals in $ ( 0,\infty ) $.
Similarly, for $ N_s = 2n + 1 $, $ \delta(x) = 1 - x R^{(n,n)}(x^2) $
is positive for $ 0 < x < 1 $, zero at $ x = 1 $, and negative for $ x > 1 $.
Thus $ \delta(x) $ has only two alternate change of
sign for any $ N_s = 2n + 1 $.
Therefore, according to de la Vall\'{e}e-Poussin's theorem
and Chebycheff's theorem, we conclude that both $ R^{(n-1,n)}(x^2) $ and
$ R^{(n,n)}(x^2) $ cannot be the optimal rational approximation
for $ (x^2)^{-1/2} $, and (\ref{eq:S_polar}) is {\it not} the
optimal rational approximation for the sign function of $ H $.
In other words, for any finite $ N_s $, the domain wall fermion
action (\ref{eq:shamir}) does {\it not} preserve the chiral symmetry optimally,
which in fact underlies the essential difficulties encountered
in lattice QCD calculations with domain wall fermions.

Note that even if one projects out the low-lying eigenmodes of $ H $
\cite{Hernandez:2000iw} (or just the boundary term of the transfer
matrix \cite{Edwards:2000qv}), treats them exactly, and
transforms $ H $ into one with narrower spectrum 
(i.e., with a smaller value of the ratio $ \lambda_{max}^2/\lambda_{min}^2 $)
such that the chiral symmetry of (\ref{eq:propagator}) is improved,
however, in principle, (\ref{eq:S_polar}) still does not satisfy the
criterion for the optimal rational approximation of the sign function
of $ H $, regardless of the spectrum of $ H $.

The optimal rational approximation for the inverse square root function
was first obtained by Zolotarev in 1877 \cite{Zol:1877}, using
Jacobian elliptic functions.
A detailed discussion of Zolotarev's result can be found in
Akhiezer's two books \cite{Akhiezer:1992,Akhiezer:1990}.
Unfortunately, Zolotarev's optimal rational approximation
has been overlooked by the numerical algebra community until
recent years.

For lattice QCD with DWF, the relevant problem is how to
construct a DWF action such that the operator $ S $ in the quark
propagator (\ref{eq:propagator}) is equal to
\BAN
\label{eq:S_op}
S = H R_Z(H^2)
\EAN
where $ R_Z(H^2) $ is the Zolotarev optimal rational approximation
for the inverse square root of $ H^2 $.  
In general, we have two options for $ R_Z $, namely,
\BAN
\label{eq:rz_nn}
R^{(n,n)}_Z(H^2) = \frac{d_0}{\lambda_{min}}
\prod_{l=1}^{n} \frac{ 1+ h^2/c_{2l} }{ 1+ h^2/c_{2l-1} }
\EAN
and
\BAN
\label{eq:rz_n1n}
R^{(n-1,n)}_Z(H^2) = \frac{d'_0}{\lambda_{min}}
\frac{ \prod_{l=1}^{n-1} ( 1+ h^2/c'_{2l} ) }
     { \prod_{l=1}^{n} ( 1+ h^2/c'_{2l-1} ) }
\EAN
where $ h^2 = H^2/\lambda_{min}^2 $, $ \lambda_{min}^2 $  
($ \lambda_{max}^2 $) is the minimum (maximum) of the eigenvalues of $ H^2 $,
and the coefficients $ d_0 $, $ d'_0 $, $ c_l $ and $ c'_l $
are expressed in terms of elliptic functions \cite{Akhiezer:1990}
with arguments depending on $ n $ and $ \lambda_{max}^2 / \lambda_{min}^2 $.

Now if one could construct a domain wall fermion action
such that the operator $ T $ in (\ref{eq:T}) is replaced with
\bea
\label{eq:T_var}
{\cal T} = \prod_{s=1}^{N_s} \frac{ 1 - \omega_s H }{ 1 + \omega_s H} \ ,
\eea
then one can solve for $ \{ \omega_s \} $
such that the operator $ S $ (\ref{eq:S}) is equal to
\bea
{\cal S}
= \frac{1-{\cal T}}{1+{\cal T}}
= \left\{ \begin{array}{ll}
           H R_Z^{(n,n)}(H^2),  &   N_s = 2n+1  \nn
           H R_Z^{(n-1,n)}(H^2), &  N_s = 2n  \nn
          \end{array} \right. \\
\label{eq:S_var}
\eea
Note that one does not have the option to put different weights for $ H $
in the numerator and the denominator of (\ref{eq:T_var}), since the optimal
rational approximation of the sign function of $ H $ is
equal to $ H $ times the optimal rational approximation of $ (H^2)^{-1/2} $.
Obviously, the highest degree $ n $ one can obtain with
$ N_s $ flavors is only $ [\frac{N_s}{2}] $.

Nevertheless, it seems to be nontrivial to implement
the weights $ \{ \omega_s \} $ into the DWF 
action (\ref{eq:shamir}) such that (\ref{eq:T_var}) can be reproduced.

Instead of working with the domain wall fermion action (\ref{eq:shamir}),
I consider one of its variants \cite{Borici:1999zw}, which differs
from (\ref{eq:shamir}) by replacing $ \delta_{x,x'} D_5(s,s') $ with
\bea
\label{eq:borici}
D_5 (x,s;x',s')
=\delta_{x,x'}\delta_{s,s'}+(D_w-1)_{x,x'} P_{-} \delta_{s',s+1}
+(D_w-1)_{x,x'} P_{+} \delta_{s',s-1}.
\eea

Then the quark propagator in a background gauge field can be
evaluated \cite{Edwards:2000qv} as
\bea
\label{eq:propagator_b}
\left< q(x) \bar{q}(y) \right>
= \frac{1-\gamma_5 S_w}{1+\gamma_5 S_w}
\eea
where $ S_w $ is same as (\ref{eq:S}) except substituting $ H $ with
$ H_w = \gamma_5 D_w $. Evidently, (\ref{eq:propagator_b})
does {\it not} preserve the chiral symmetry optimally,
the argument is same as the case of (\ref{eq:propagator}).

In view of (\ref{eq:T_var}) and (\ref{eq:S_var}), now it is clear 
how to construct the optimal domain wall fermion action on the lattice.
Explicitly, it reads
\bea
\label{eq:twc}
&& \hspace{-6mm}
{\cal A} = \sum_{s,s'=1}^{N_s} \sum_{x,x'}
\bar\psi(x,s)
 [ ( 1 + \omega_s D_w )_{x,x'} \delta_{s,s'}
 - (1-\omega_{s} D_w)_{x,x'} P_{-} \delta_{s',s+1} \nn
&& \hspace{50mm}
 - (1-\omega_{s} D_w)_{x,x'} P_{+} \delta_{s',s-1} ] \psi(x',s')
\eea
with weights
\bea
\label{eq:omega}
\omega_s = \frac{1}{\lambda_{min}} \sqrt{ 1 - \kappa'^2 \mbox{sn}^2
                  \left( v_s ; \kappa' \right) } \ ,
\eea
where $ \mbox{sn}( v_s; \kappa' ) $ is the Jacobian elliptic function
with argument $ v_s $ (\ref{eq:vs}) and modulus
$ \kappa' = \sqrt{ 1 - \lambda_{min}^2 / \lambda_{max}^2 } $
($ \lambda_{max}^2 $ and $ \lambda_{min}^2 $ are the maximum
and the minimum of the eigenvalues of $ H_w^2 $), and
$ \{ \omega_s \} $ are obtained from the roots
$ (u_s = \omega_s^{-2}, s=1,\cdots,N_s) $ of the equation
\BAN
\label{eq:delta_Z}
\delta_Z(u) =
     \left\{ \begin{array}{ll}
 1-\sqrt{u} R_Z^{(n,n)}(u)=0 \ ,  &  \ N_s=2n+1 \nn
 1-\sqrt{u} R_Z^{(n-1,n)}(u)=0 \ , & \ N_s=2n   \nn
             \end{array} \right.
\EAN
It can be shown that the argument $ v_s $ in (\ref{eq:omega})
is
\bea
\label{eq:vs}
v_s &=& (-1)^{s-1} M \ \mbox{sn}^{-1}
    \left( \sqrt{\frac{1+3\lambda}{(1+\lambda)^3}}; \sqrt{1-\lambda^2} \right)
  + \left[ \frac{s}{2} \right] \frac{2K'}{N_s}
\eea
where
\bea
\label{eq:lambda}
\lambda &=&
\prod_{l=1}^{N_s}
\frac{\Theta^2 \left(\frac{2lK'}{N_s};\kappa' \right)}
     {\Theta^2 \left(\frac{(2l-1)K'}{N_s};\kappa' \right)} \ , \\
\label{eq:M}
M &=&
\prod_{l=1}^{[\frac{N_s}{2}]}
\frac{\mbox{sn}^2 \left(\frac{(2l-1)K'}{N_s};\kappa' \right) }
{ \mbox{sn}^2 \left(\frac{2lK'}{N_s};\kappa' \right) } \ ,
\eea
$ K' $ is the complete elliptic integral of the first kind with 
modulus $ \kappa' $, and $ \Theta $ is the elliptic theta function.
From (\ref{eq:omega}), it is clear that
$ \lambda_{max}^{-1} \le \omega_s \le \lambda_{min}^{-1} $ 
since $ \mbox{sn}^2(;) \le 1 $. 

The quark propagator in a background gauge field can be derived as
\bea
\label{eq:propagator_opt}
\left< q(x) \bar{q}(y) \right>
= \frac{1-\gamma_5 {\cal S}_{opt}}{1+\gamma_5 {\cal S}_{opt}}
\eea
where $ {\cal S}_{opt} $ is same as (\ref{eq:S_var}) except 
substituting $ H $ with $ H_w $.

Since the chiral symmetry of (\ref{eq:propagator_opt}) is equivalent
to $ {\cal S}_{opt}^2 = 1 $, its breaking due to a finite $ N_s $ can be
measured in terms of the deviation
\BAN
\label{eq:Delta_Z}
\Delta_Z = \max_{ \forall \ Y \ne 0 }
\left|\frac{Y^{\dagger} {\cal S}_{opt}^2 Y}{Y^{\dagger}Y}-1 \right| \ ,
\EAN
which has a theoretical upper bound \cite{Chiu:2002eh},
$ 2 ( 1 - \lambda )/(1+\lambda) $, where $ \lambda $ is defined in
(\ref{eq:lambda}), a function
of $ N_s $ and $ b = \lambda_{max}^2 / \lambda_{min}^2 $.
In practice, with $ N_s = 32 $, one should have no difficulties to
achieve $ \Delta_Z < 10^{-12} $ for any gauge configurations
on a finite lattice (say, $ 16^3 \times 32 $ at $ \beta = 6.0 $).

It is simple to incorporate the bare quark mass $ m_q $
by adding the following terms 
\BAN
\label{eq:qmass}
\hspace{-8mm}
&& \frac{m_q}{2m_0} \sum_{x,x'} \
     [ \bar\psi(x,1) (1-\omega_{1} D_w)_{x,x'} P_{+} \psi(x',N_s) \nn
&& \hspace{15mm}
     + \bar\psi(x,N_s) (1-\omega_{N_s} D_w)_{x,x'} P_{-} \psi(x',1) ],
\EAN
to the optimal DWF action (\ref{eq:twc}), and changing
the boundary conditions to
\BAN
P_{+} \psi(x,0) = - \frac{m_q}{2m_0} P_{+} \psi(x,N_s), \hspace{6mm}
P_{-} \psi(x,N_s+1) = - \frac{m_q}{2m_0} P_{-} \psi(x,1) \ .
\label{eq:bc}
\EAN

After introducing pseudofermions (Pauli-Villars fields)
with fixed mass $ m_q = 2 m_0 $, one can derive
the effective 4D lattice Dirac operator for the internal
quark loops as
\bea
\label{eq:D}
D(m_q) = r( D_c + m_q )( 1 + r D_c )^{-1} \ ,
\hspace{4mm} r = \frac{1}{2m_0}, 
\eea
where $ r D_c $ denotes the inverse of the 
massless quark propagator (\ref{eq:propagator_opt}) which
becomes chirally symmetric in the limit $ N_s \to \infty $.
The exponential locality of $ D $ (\ref{eq:D}) (for any $ m_q $ and $ N_s $)
has been asserted for sufficiently smooth gauge background \cite{Chiu:2002kj}.

In the massless limit ($m_q=0$) and $ N_s \to \infty $,
$ D $ (\ref{eq:D}) is exactly equal to the overlap Dirac operator 
\cite{Neuberger:1998fp,Narayanan:1995gw}, and  
satisfies the Ginsparg-Wilson relation \cite{Ginsparg:1981bj}
\BAN
D \gamma_5 + \gamma_5 D = 2 D \gamma_5 D \ .
\EAN
This implies that $ D $ is topologically-proper
(i.e., with the correct index and axial anomaly),
similar to the case of overlap Dirac operator.
For any finite $ N_s $, $ D $ is exactly equal to the overlap Dirac
operator with $ (H_w)^{-1/2} $ approximated by Zolotarev rational polynomial.

From (\ref{eq:D}), the valence quark propagator coupling to physical
hadrons can be expressed as
\BAN
\label{eq:valence_q}
( D_c + m_q )^{-1} = r ( 1 - r m_q )^{-1} [ D^{-1}(m_q) - 1 ] \ ,
\EAN
where $ D^{-1}(m_q) $ can be computed via the 5-dimensional
lattice Dirac operator of optimal DWF. Evidently, the
valence quark propagator of optimal DWF is exactly equal 
to that of the overlap with Zolotarev approximation.   
Preliminary numerical results have demonstrated that the quark
propagator of optimal DWF with $ N_s = 2n $ is precisely  
equal\footnote{The relative error between these two quark propagators
is always less than $ 10^{-7} $ for stopping criterion $ 10^{-11} $ in
the conjugate gradient loops.} to that of the overlap
with $ (H_w^2)^{-1/2} $ approximated by $ R_Z^{(n-1,n)}(H_w^2) $. 
 
A simple way to improve the chiral symmetry of DWF action (\ref{eq:shamir})
is to replace $ D_w $ with $ \omega_s D_w $,
\bea
\label{eq:twc1}
{\cal A}'_{dwf} = \sum_{s,s'=1}^{N_s} \sum_{x,x'}
\bar\psi(x,s)
[ \omega_s D_w(x,x') \delta_{s,s'} + \delta_{x,x'}  D_5(s,s') ] \psi(x',s')
\eea
where $ \{ \omega_s \} $ are given in (\ref{eq:omega}).
It is easy to see that in the limit $ a_5 \to 0 $, 
both (\ref{eq:twc}) and (\ref{eq:twc1}) give the
same quark propagator (\ref{eq:propagator_opt})
with optimal chiral symmetry.
In Table 1, the precision of chiral symmetry of each DWF action
(with $ N_s = 16 $) discussed above is measured in terms of
$ \sigma \equiv
\max_{i,j} \left| 
({\cal D}^{-1} \gamma_5 + \gamma_5 {\cal D}^{-1} )_{ij} \right|
$, where $ {\cal D}^{-1} = \langle q \bar{q} \rangle $ is
the quenched {\it massless} quark propagator  
(with one of its endpoints fixed at origin)
in a gauge background generated with Wilson gauge action at
$ \beta = 6.0 $ on the $ 8^4 $ lattice, and the ranges of the
indices are: $ 1 \le i \le 12 \times 8^4 $, $ 1 \le j \le 12 $.
The eigenvalues of $ H_w $ are bounded as
$ | \lambda(H_w) | \in  [0.1848, 6.5348] $ for $ m_0 = 1.0 $, 
while $ | \lambda(H_w) | \in  [0.0946, 5.7484] $ for $ m_0 = 1.8 $.
The quark propagators
are computed by conjugate gradient with stopping criterion $ 10^{-11} $.
Evidently, the improved DWF action (\ref{eq:twc1}) preserves
the chiral symmetry much better than
Shamir's action and Borici's variant,
and the optimal DWF (\ref{eq:twc}) is the best among these DWF actions.
Finally, we note that (\ref{eq:twc1}) can be easily implemented
for machineries already geared to (\ref{eq:shamir}).

{\footnotesize
\begin{table}
\begin{center}
\begin{tabular}{|c|c|c|}
\hline
DWF action & $ \sigma (m_0=1.0) $ & $ \sigma (m_0=1.8) $  \\
\hline
\hline
Shamir (\ref{eq:shamir}) & $ 4.3 \times 10^{-5} $ & $ 1.9 \times 10^{-5} $ \\
\hline
Borici (\ref{eq:shamir}) \& (\ref{eq:borici}) & $ 5.4 \times 10^{-4} $ &
$ 1.0 \times 10^{-4} $ \\
\hline
Improved DWF (\ref{eq:twc1}) & $ 2.4 \times 10^{-6} $ &
$ 3.5 \times 10^{-8} $ \\
\hline
Optimal DWF (\ref{eq:twc}) & $ 8.8 \times 10^{-9} $ & $3.8 \times 10^{-10}$ \\
\hline
\end{tabular}
\end{center}
\caption{The precision of chiral symmetry of the massless 
quark propagator in a gauge background on the 
$ 8^4 $ lattice at $ \beta = 6.0 $,
for various domain-wall fermion actions with $N_s = 16$.
(Note that in the limit $ N_s \to \infty $,
$ \sigma \to 0 $ for all DWF actions.) }
\label{table:sigma}
\end{table}
}

In summary, the problem how to construct a DWF action such that
the effective 4D lattice Dirac operator can preserve the
chiral symmetry optimally for any given finite $ N_s $,
has been solved in (\ref{eq:twc}).
It provides a better understanding of exact chiral
symmetry on a finite lattice, as well as the optimal way to
tackle nonperturbative issues in QCD or supersymmetric QFTs.



This work was supported in part by National Science Council,
ROC, under the grant number NSC91-2112-M002-025, and NSC-40004F.
I would like to thank Steve Sharpe, David Kaplan,
and the Particle Theory Group at University of Washington
for their kind hospitality. The numerical results in Table~1  
were computed with the Linux PC cluster (hep-lat/0208039) at NTU,
and I thank Tung-Han Hsieh for his help in performing the tests.

\bigskip
\bigskip

\vfill\eject


\begin{thebibliography}{15}

\bibitem{Rubakov:bb}
V.~A.~Rubakov and M.~E.~Shaposhnikov,
Phys.\ Lett.\ B {\bf 125}, 136 (1983).

\bibitem{Callan:sa}
C.~G.~Callan and J.~A.~Harvey,
Nucl.\ Phys.\ B {\bf 250}, 427 (1985).


\bibitem{Kaplan:1992bt}
D.~B.~Kaplan,
Phys.\ Lett.\ B {\bf 288}, 342 (1992)


\bibitem{Shamir:1993zy}
Y.~Shamir,
Nucl.\ Phys.\ B {\bf 406}, 90 (1993)


\bibitem{Neuberger:1997bg}
H.~Neuberger,
Phys.\ Rev.\ D {\bf 57}, 5417 (1998)


\bibitem{Kikukawa:1999sy}
Y.~Kikukawa and T.~Noguchi,
hep-lat/9902022.

\bibitem{Akhiezer:1992}
N.~I.~Akhiezer,
"Theory of approximation", Reprint of 1956 English translation, Dover,
New York, 1992.

\bibitem{Hernandez:2000iw}
P.~Hernandez, K.~Jansen and M.~Luscher,
hep-lat/0007015.

\bibitem{Edwards:2000qv}
R.~G.~Edwards and U.~M.~Heller,
Phys.\ Rev.\ D {\bf 63}, 094505 (2001)


\bibitem{Zol:1877}
E.~I.~Zolotarev,
Zap. Imp. Akad. Nauk. St. Petersburg, 30 (1877), no. 5; reprinted in his
Collected works, Vol. 2, Izdat, Akad. Nauk SSSR, Moscow, 1932, p. 1-59.


\bibitem{Akhiezer:1990}
N.~I.~Akhiezer,
"Elements of the theory of elliptic functions",
Translations of Mathematical Monographs, 79,
American Mathematical Society, Providence, R.I. 1990.


\bibitem{Borici:1999zw}
A.~Borici,
Nucl.\ Phys.\ Proc.\ Suppl.\  {\bf 83}, 771 (2000)


\bibitem{Chiu:2002eh}
T.~W.~Chiu, T.~H.~Hsieh, C.~H.~Huang and T.~R.~Huang,
Phys.\ Rev.\ D {\bf 66}, 114502 (2002)

\bibitem{Neuberger:1998fp}
H.~Neuberger,
Phys.\ Lett.\ B {\bf 417}, 141 (1998);
%
Phys.\ Lett.\ B {\bf 427}, 353 (1998)

\bibitem{Narayanan:1995gw}
R.~Narayanan and H.~Neuberger,
Nucl.\ Phys.\ B {\bf 443}, 305 (1995)

\bibitem{Ginsparg:1981bj}
P.~H.~Ginsparg and K.~G.~Wilson,
Phys.\ Rev.\ D {\bf 25}, 2649 (1982)


\bibitem{Chiu:2002kj}
T.~W.~Chiu,
Phys.\ Lett.\ B {\bf 552}, 97 (2003)


\end{thebibliography}
\end{document}